# The role of Cr substitution on the ferromagnetic properties of $Ga_{1-x}Cr_xN$

R.K. Singh,[a] Stephen Y. Wu,[a] H.X. Liu,[a] Lin Gu,[b] David J. Smith,[b),c] and N. Newman[a),b),d]

[a] Department of Chemical & Materials Engineering, Arizona State University, Tempe, Arizona 85287-6006

[b] Center for Solid State Science, Arizona State University, Tempe, Arizona 85287-1704

[c] Department of Physics & Astronomy, Arizona State University, Tempe, Arizona 85287-1504

Angular-dependent channeling Rutherford Backscattering Spectroscopy (c-RBS) has been used to quantify the fraction of Cr atoms on substitutional, interstitial, and random sites in epitaxial $Ga_{1-x}Cr_xN$ films grown by reactive molecular-beam epitaxy. The morphology of these films and correlation with their magnetic properties has been investigated. Films grown at temperatures below ~ 750°C have up to 90% of Cr occupying substitutional sites. Post-growth annealing at 825°C results in a systematic drop in the fraction of substitutional Cr as well as a fall off in the ferromagnetic signal. The roles of non-substitutional Cr in transferring charge from the Cr t2 band and segregation of substitutional Cr in the loss of magnetism are discussed. Overall, these results provide strong microscopic evidence that Cr-doped III-N systems are dilute magnetic semiconductors.

[d] electronic mail: Nathan.Newman@asu.edu

Dilute magnetic semiconductors (DMSs) with Curie temperature ($T_c$) significantly above room temperature are needed for the development of practical spin-dependent electronic devices. Many groups have demonstrated room temperature ferromagnetism in a variety of wide band gap doped semiconductors, such as $TiO_2$,[1-3] ZnO,[4-6] $SnO_2$,[7] $ZnGeP_2$,[8] $CdGeP_2$,[9] $CuGaSe_2$,[10] AlN,[11-14] and GaN.[13-16] We have earlier reported ferromagnetism with $T_c$ above 900 K (~ 625°C) in both $Ga_{1-x}Cr_xN$ and $Al_{1-x}Cr_xN$.[13,14] However, despite recent experimental successes, a fundamental understanding of the origin of ferromagnetism in these DMSs is incomplete. There is still controversy whether secondary phases or other forms of inhomogeneities are responsible for the ferromagnetic ordering, and whether these materials are truly carrier-mediated. There are reports indicating that non-optimal synthesis conditions lead to the formation of secondary phases that are responsible for spin-glass behavior[17] or high temperature ferromagnetic ordering.[18]

A theoretical study using first-principles calculations shows that transition metal-doped GaN is a promising candidate as a room temperature DMS and that Cr-doped GaN has the most stable ferromagnetic states.[19] Since the Cr $t_2$ defect level is expected to be 1/3-filled, the partial compensation from the commonly observed background donors that make GaN naturally n-type will drive the system towards optimal 1/2-filling for the double exchange mechanism of ferromagnetism. The lower vapor pressure of Cr also allows the growth of high quality Cr-doped GaN films. High volatility of Mn, on the other hand, limits the growth temperature for deposition of Mn-doped III-V DMSs.[20]

It is well established that 3d impurities in GaAs and other conventional III-V semiconductors are largely substitutional for small to moderate doping concentrations.[21] It has, however, been recently shown that Mn impurity atoms at high concentrations occupy interstitial sites and degrade the magnetic properties of the films.[20,21] These studies clearly indicate that site preference is an important factor in determining the ferromagnetic behavior of these doped semiconductors.

There is currently lack of information about the location of Cr atoms in the GaN and AlN lattice, as well as the magnetic state of these ions. Cr atoms can occupy three types of sites in the $Ga_{1-x}Cr_xN$ lattice: (1) substitutional Ga lattice sites to form a $Ga_{1-x}Cr_xN$ alloy; (2) interstitial sites commensurate with the wurzite lattice structure; and (3) random sites, which are incommensurate with the wurzite lattice, by segregating or precipitating out as different phases (e.g., $Cr_xN$ inclusions). In this work we report quantitative determinations of the lattice site occupancy for as-deposited and annealed Cr-doped GaN epitaxial films as determined by ion channeling Rutherford Backscattering Spectrometry (c-RBS), and nuclear reaction analysis (NRA). These studies establish strong correlation between the arrangement of Cr in $Ga_{1-x}Cr_xN$ and its ferromagnetic behavior.

The epitaxial Cr-GaN films studied in these experiments were ~ 200 nm thick, and were grown on c-axis 6H-SiC (0001) substrates by reactive molecular beam epitaxy under a typical base pressure of $< 5x10^{-10}$ Torr. The Cr concentration in all films was ~ 3 at. %, which is expected to give the maximum magnetic moment in this system.[14] The thickness and depth profile of the chemical composition were characterized by RBS. A 2 MeV $He^{++}$ beam was used to obtain channeling yields of Ga and Cr, and a 3.73 MeV $He^{++}$ beam was used to obtain channeling yields of N by NRA. Nuclear resonant elastic scattering, $^{14}N$ $(\alpha,\alpha)$ $^{14}N$ was used for N detection. The channeling angular distributions for Cr and Ga were obtained simultaneously using a two-axis goniometer. The channeling angular distributions of N were obtained in the same way using NRA. Structural properties were characterized using X-ray diffraction (XRD) [Rigaku D/MAX-IIB with a single crystal graphite monochromator], high resolution XRD (HRXRD) [Bede D1 system with dual channel single crystal Si monochromator], and transmission electron microscopy (TEM) [JEOL 4000EX]. The magnetic properties were characterized over the temperature range from -260°C to 825°C with a Quantum Design vibrating sample magnetometer (VSM) equipped with the recently developed oven option for the Physical Property Measurement System (PPMS). Magnetic fields were applied parallel to the film plane during susceptibility measurements. The analysis of temperature dependent magnetization at 0.5 T did not show evidence of additional paramagnetic moment from non-ferromagnetic Cr. Additional details of the growth and characterization experiments are given elsewhere.[12]

Cr-GaN films grown at temperatures above 700°C provided excellent quality epitaxial GaN layers, as indicated by a 2 MeV He ion channeling minimum yield ($\chi_{min}$) of better than 3%. High backscattering yield ($\chi_{min}$ = 12%) observed near the substrate-film interface suggests that some strain or dislocation defects are introduced during initial growth. Structural characterization by HRTEM confirms the presence of higher defect density near the substrate-film interface. A similar trend is observed in all the films. Additional information about the structure of the Cr-GaN films is given in Fig. 1. As expected, the FWHM decreased with increasing growth temperature. The lattice constant, on the other hand, is minimized for the 775°C growth temperature. The change in lattice constant can be correlated to the fraction of substitutional Cr on the Ga site. Extensive TEM observation and XRD scans of samples grown at 700°C, 740°C, and 775°C did not provide evidence for any secondary phases (CrN, $Cr_2N$, $CrGa_4$ or $CrO_2$), although small isolated patches of cubic GaN were occasionally observed close to the substrate.

Figure 2 shows the channeling angular distributions of Cr and Ga in the <0001> axial direction. The angular distributions of Cr for films grown at 700°C, 740°C, and 775°C exhibit small normalized yields ($\chi_{Cr}$) compared to the normalized yield of Cr for film grown at 825°C.

The fraction of Cr in substitutional site ($Cr_{Ga}$) of the $Ga_{1-x}Cr_xN$ films was calculated from the measured $\chi_{Ga}$ and $\chi_{Cr}$ using the expression,[15] $Cr_{Ga} = (1 - \chi_{Cr}) / (1 - \chi_{Ga})$. The results indicate that 78 to 90 % of Cr occupies substitutional sites for GaN films grown at temperatures ≤ 775°C, depending on the particular growth temperature. However, only a small fraction of Cr (< 20%) is located in substitutional sites for film grown at 825°C. HRTEM results indicate that films grown at ≤ 775°C have uniform Cr distribution in the lattice (Fig. 3(a)), while energy-filtered TEM imaging of films grown at 825°C, with low substitutional Cr, shows significant Cr clustering (Fig. 3(b)). Line profile analysis of Cr with a spatial resolution of ~ 1 nm was performed using EELS (Electron Energy Loss Spectroscopy). The results indicate that the Cr concentration is homogeneous for the films grown at ≤ 775°C, whereas considerable undulation, with a length scale of a few hundred Ångstroms, is observed for the film grown at 825°C (Insets of Figs. 3(a) and (b)).

The measured magnetic susceptibility in these films synthesized with identical Cr concentrations (3%) varies significantly with the fraction of substitutional Cr in the GaN lattice. Although we have observed a magnetic moment of up to 1.8 $\mu_B$/Cr, indicating that 60% of the Cr is magnetically active, in films with high substitutional Cr (>80%), we notice a gradual decay in the magnetic moment to ~ 20% magnetically active Cr over a period of time. In this investigation, we study the influence of annealing on films that exhibit a stable signal with storage time and measurement.

The Cr-GaN film grown at 775°C has the highest saturation magnetic moment ($M_s$) of 0.35 $\mu_B$/Cr, indicating that ~ 12% of Cr is magnetically active (Fig. 4). For the films grown at 825°C, with high random Cr content (>80%), the saturation magnetic moment, and the fraction of magnetically active Cr in the film, drop to 0.13 and 4%, respectively. These results give indisputable evidence that substitutional Cr is strongly involved in the observed ferromagnetism of these films. Although the difference between the amount of random Cr in the films grown at 700°C, 740°C, and 775°C is close to the experimental error, we observe much higher saturation magnetic moment for the film grown at 775°C. This could be due to smaller lattice constant and better crystallinity obtained in this film (Fig. 1). Further study to evaluate this possibility is in progress.

Channeling angular scans performed on the films after annealing at 825°C show a systematic drop in the fraction of substitutional Cr and the corresponding ferromagnetic signal (Fig. 4). This behavior establishes a direct correlation between the microscopic Cr on Ga site ($Cr_{Ga}$) defect and the magnetic properties. The double exchange mechanism relies on establishing long-range ferromagnetic order by localized hopping within the $t_2$ Cr defect band. We speculate that the rapid decrease in effective saturation magnetic moment per Cr atom at higher growth temperatures (>775°C) and after annealing at 825°C may be attributed to enhanced antiferromagnetic coupling as a result of some combination of four factors: (1) decreased fraction of Cr on the substitutional site; (2) increase in the inhomogeneity of the substitutional Cr distribution; (3) non-substitutional Cr coupling antiferromagnetically with the remaining substitutional Cr; and/or (4) change in compensation of the Cr $t_2$ band as a result of transfer of charge to other defects and/or secondary phases, if present. These factors may also account for why the processing-induced changes in magnetization are greater than that of the substitutional Cr fraction.

In conclusion, we have demonstrated that high substitution fractions of Cr in GaN can be obtained by controlled growth temperatures (≤ 775°C). The large variations in the magnetic properties of $Ga_{1-x}Cr_xN$ films grown at different temperatures and also as a result of high temperature annealing can be attributed to the lattice site rearrangement of Cr atoms. Our results establish that the location of Cr sites in the $Ga_{1-x}Cr_xN$ lattice plays a crucial role in determining its magnetic properties. The concentration of uncompensated Cr spins is controlled by the behavior of unstable defects involving mobile Cr atoms. Extensive structural characterization using XRD and TEM allows us to exclude the possibility of secondary phases or other extraneous factors in contributing to the observed magnetic behavior.

This work was supported by the Defense Advanced Research Projects Agency (DARPA) and administered by the Office of Naval Research (Contract No. N00014-02-1-0598). We acknowledge use of facilities in the Center for Solid State Science at ASU. The authors thank Barry Wilkins for his assistance with angular scans. We also thank Art Freeman and Mark van Schilfgaarde for enlightening discussions.

[1]Y. Matsumoto, M. Murakami, T. Shono, T. Hasegawa, T. Fukumora, M. Kawasaki, P. Ahmet, T. Chikyow, S. Koshihara, and H. Koinuma, Science **291**, 854 (2001).

[2]Z. Wang, W. Wang, J. Tang, L.D. Tung, L. Spinu, and W. Zhou, Appl. Phys. Lett. **83**, 518 (2003).

[3]N. H. Hong, J. Sakai, and A. Hassini, Appl. Phys. Lett. **84**, 2602 (2004).

[4]K. Ueda, H. Tabata, and T. Kawai, Appl. Phys. Lett. **79**, 988 (2001).

[5]H. Saeki, H. Tabata, and T. Kawai, Solid State Commun. **120**, 439 (2001).

[6]Y.M. Cho, W.K. Choo, H. Kim, D. Kim, and Y. Ihm, Appl. Phys. Lett. **80**, 3358 (2002).

[7]S.B. Ogale, R.J. Choudhary, J.P. Buban, S.E. Lofland, S.R. Shinde, S.N. Kale, V.N. Kulkarni, J. Higgins, C. Lanci, J.R. Simpson, N.D. Browning, S. Das Sarma, H.D. Drew, R.L. Greene, and T. Venkatesan, Phys. Rev. Lett. **91**, 077205 (2003).

[8]S. Cho, S. Choi, G.B. Cha, S.C. Hong, Y. Kim, Y.J. Zhao, A.J. Freeman, J.B. Ketterson, B.J. Kim, Y.C. Kim, and B.C. Choi, Phys. Rev. Lett. **88**, 257203-1 (2002).

[9]G.A. Medvedkin, T. Ishibashi, T. Nishi, and K. Hiyata, Jap. J. Appl. Phys. **39**, L949 (2000).

[10]Y. Zhao, P. Mahadevan, and A. Zunger, Appl. Phys. Lett. **84**, 3753 (2004).

[11]S.G. Yang, A.B. Pakhonmov, S.T. Hung, and C.Y. Wong, Appl. Phys. Lett. **81**, 2418 (2002).

[12]Stephen Y. Wu, H.X. Liu, L. Gu, R.K. Singh, L. Budd, M. van Schilfgaarde, M.R. McCartney, D.J. Smith, and N. Newman, Appl. Phys. Lett. **82**, 3047 (2003).

[13]S.Y. Wu, H.X. Liu, L. Gu, R.K. Singh, M. van Schilfgaarde, D.J. Smith, N.R. Dilley, L. Montes, M.B. Simmonds, and N. Newman, Mat. Res. Soc. Symp. Proc. **Y 10.57.1**, 798 (2004).

[14]H.X. Liu, Stephen Y. Wu, R.K. Singh, L. Gu, D.J. Smith, N.R. Dilley, L. Montes, M.B. Simmonds, and N. Newman, Appl. Phys. Lett. (to be published, Ms. # L04-0478).

[15]G.T. Thaler, M.E. Overberg, B. Gila, R. Frazier, C.R. Abernathy, S.J. Pearton, J.S. Lee, S.Y. Lee, Y.D. Park, Z.G. Khim, J. Kim, and F. Ren, Appl. Phys. Lett. **80**, 3964 (2002).

[16]M. Hashimoto, Y.K. Zhou, M. Kanamura, and H. Asahi, Solid State Commun. **37**, 122 (2002).

[17]S. Kolesnik, B. Dabrowski, and J. Mais, J. Supercond. **15**, 251 (2002).

[18]D.H. Kim, J.S. Yang, K.W. Lee, S.D. Bu, T.W. Noh, S.J. Oh, Y.W. Kim, J.S. Chung, H. Tanaka, H.Y. Lee, and T. Kawai, Appl. Phys. Lett. **81**, 2421 (2002).

[19]K. Sato and H. Katayama-Yoshida, Jpn. J. Appl. Phys. Part 2 **40**, L485 (2001).

[20]K.M. Yu, W. Walukiewicz, T. Wojtowicz, I. Kuryliszyn, X. Liu, Y. Sasaki, and J.K. Furdyna, Phys. Rev. B **65**, 201303(R) (2002).

[21]A. Zunger in Solid State Physics, edited by F. Seitz, H. Ehrenreich, and D. Turnbull, Academic Press, New York, **39**, 275 (1986).

[22]L.C. Feldman, J.W. Mayer, and S.T. Picraux, Material Analysis by Ion Channeling (Academic, New York 1982).

Figure Captions:

FIG. 1 Variation in c-plane lattice constant and rocking curve full-width-half-maximum (FWHM) as a function of growth temperature.

FIG. 2 Channeling angular scans in <0001> axial direction for Cr-GaN films grown at (a) $700^{o}$C, (b) $740^{o}$C, (c) $775^{o}$C, and (d) $825^{o}$C.

FIG. 3 (a) HRTEM image and (b) energy-filtered TEM image showing Cr distribution in GaN film grown at $775^{o}$C and $825^{o}$C respectively. EELS line profile analysis of Cr is shown as inset.

FIG. 4 Change in substitutional Cr (top) and magnetization (bottom) after annealing.

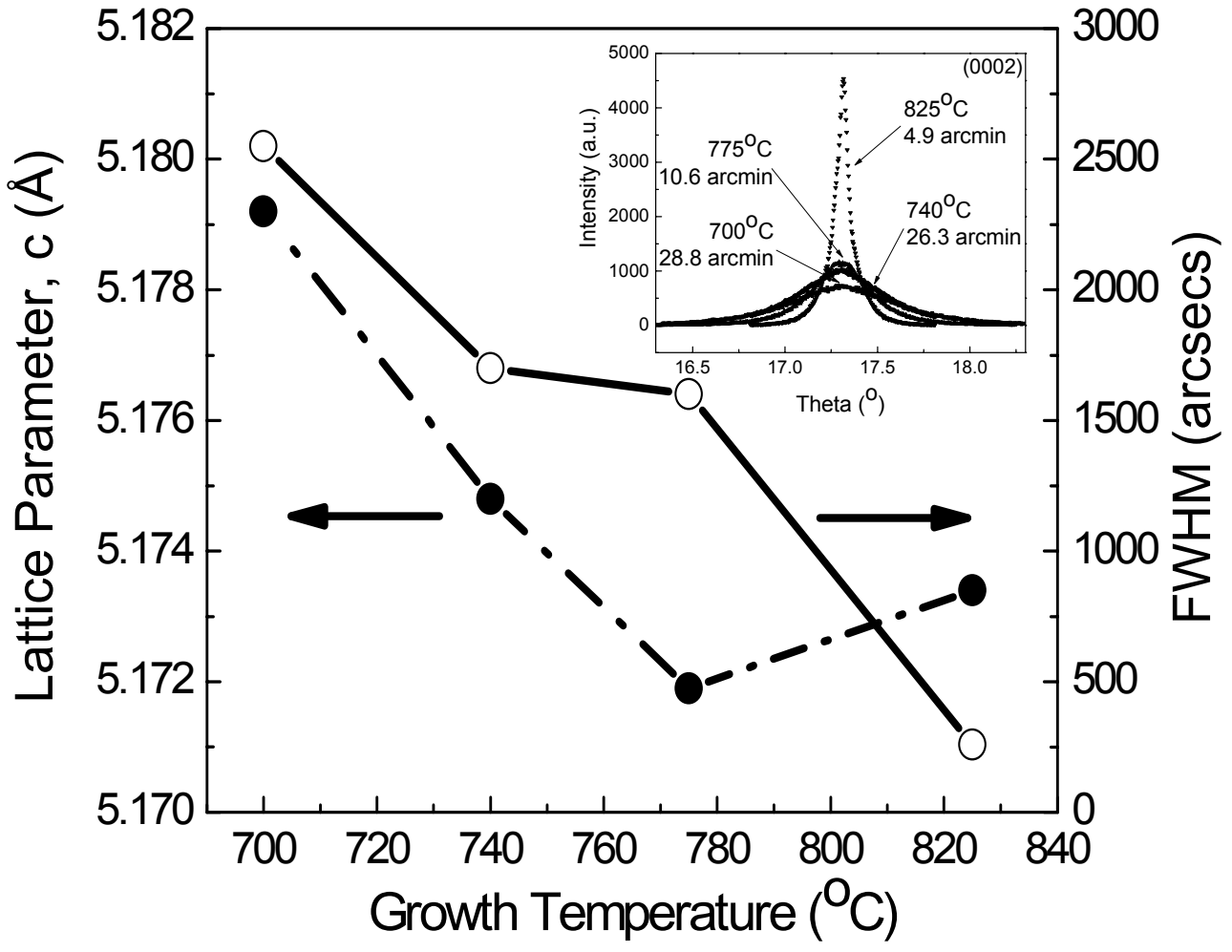

**FIG. 1 – Singh et al.**

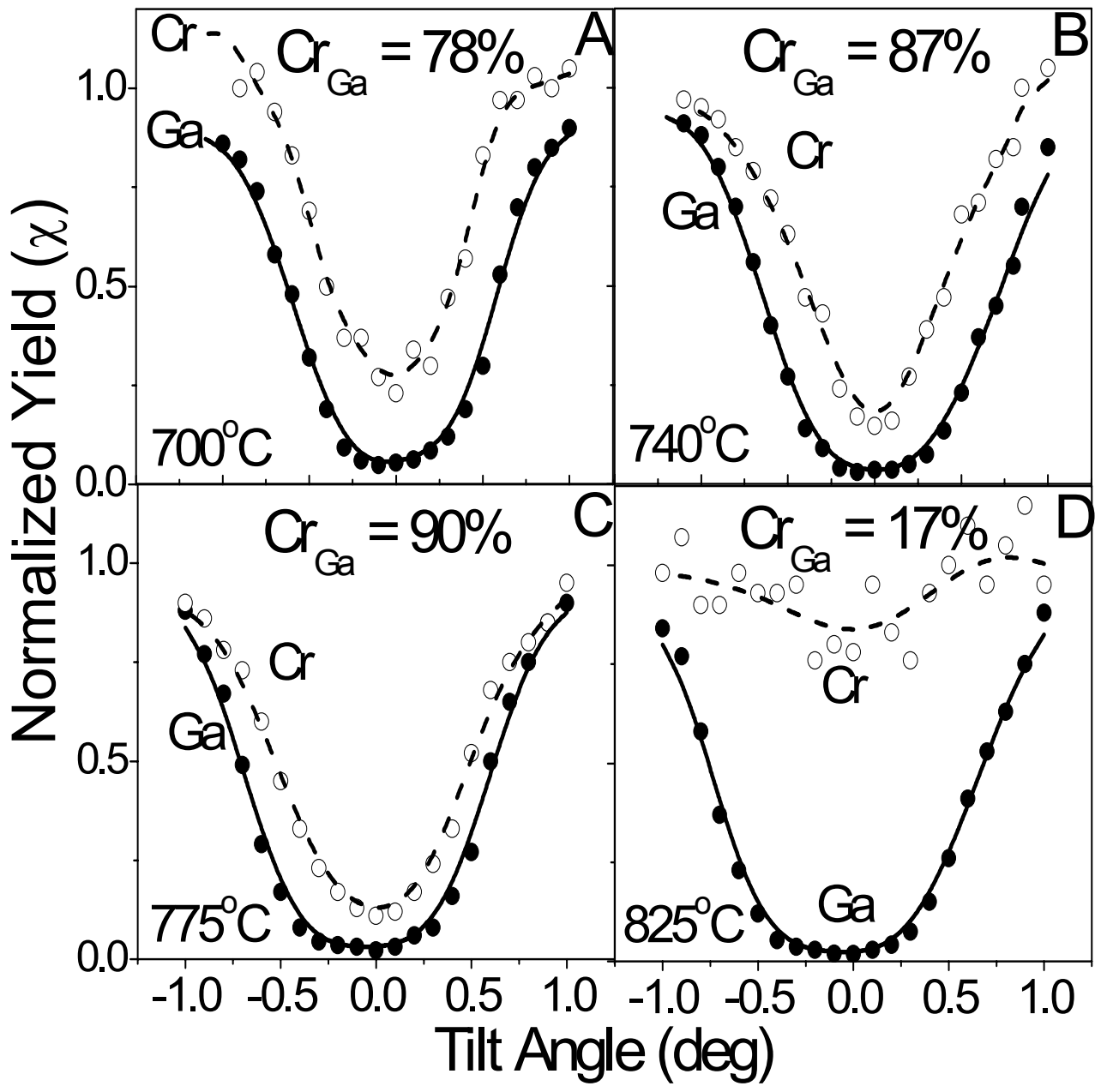

**FIG. 2 – Singh et al.**

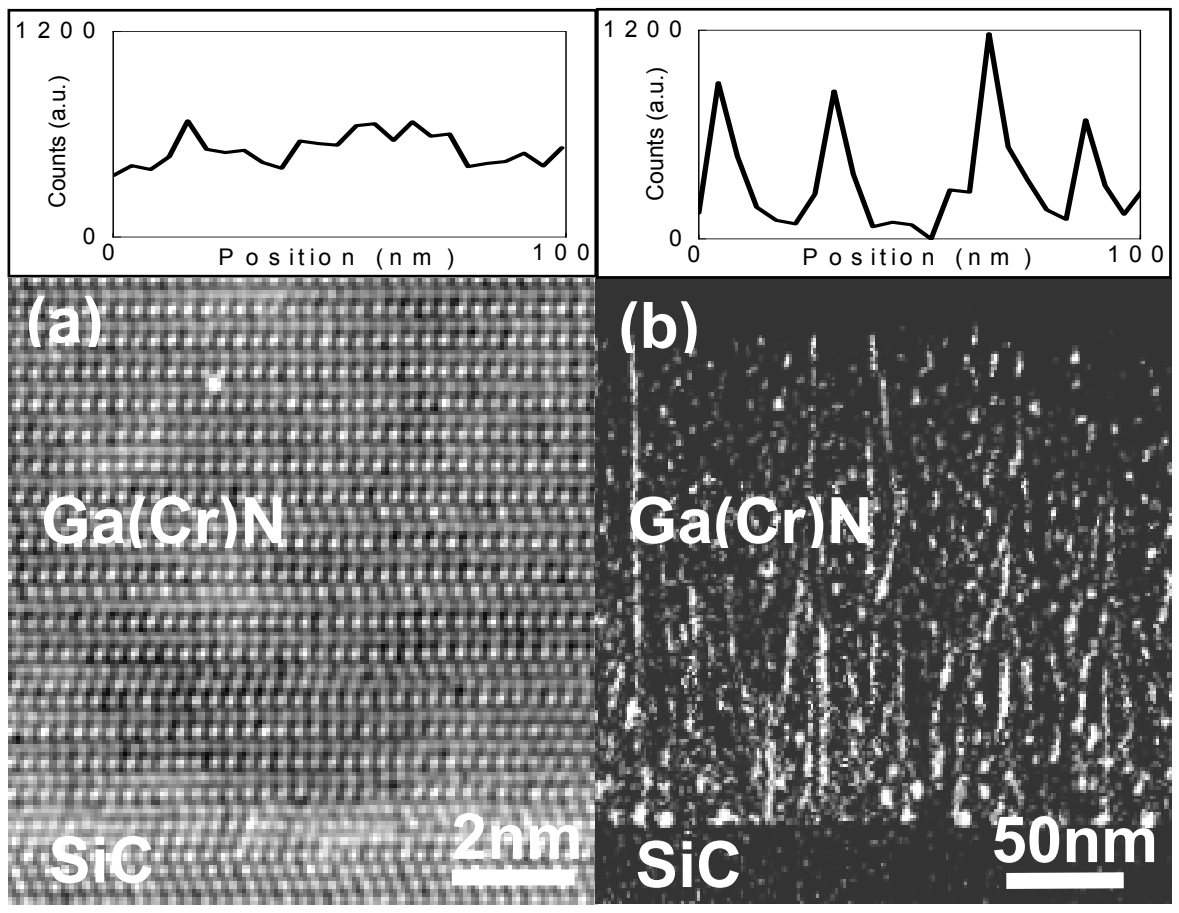

**FIG. 3 – Singh et al.**

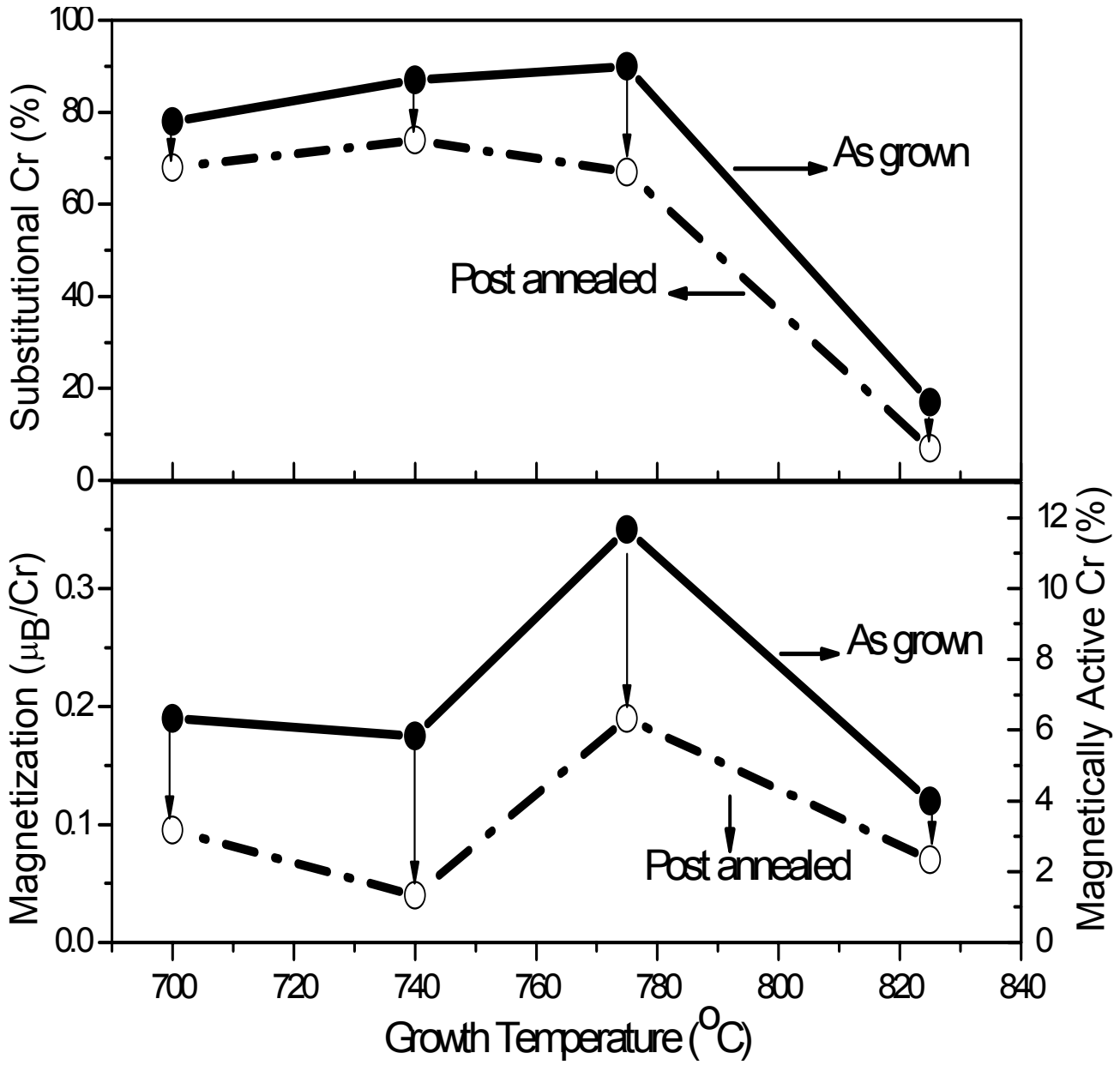

**FIG. 4 – Singh et al.**